\documentclass[pra,floatfix,preprint,nofootinbib,eqsecnum,12pt]{revtex4}
\usepackage{epsfig}
\usepackage{bm}
\usepackage{amsmath}
\usepackage{setspace}
\input epsf
\numberwithin{equation}{section}

\def\HP{$(H,\Psi)$}

\def\H{H}
\def\Dobs{D_{\text obs}}

\def\be{\begin{equation}}
\def\ee{\end{equation}}

\begin{document}
\vspace{1cm}

\title{The Impact of Cosmology on Quantum Mechanics\footnote{A pedagogical essay. Based in part on a talk given at the conference {\it 90 Years of Quantum Mechanics}. Singapore, January, 2017}}
\author{James B.~Hartle}

\email{hartle@physics.ucsb.edu}

\affiliation{Department of Physics, University of California,Santa Barbara, CA 93106-9530}
\affiliation{Santa Fe Institute, Santa Fe, NM 87501}

\date{\today }

\begin{abstract}
\singlespacing
When quantum mechanics was developed in the '20s of the last century another revolution in physics was just starting. It began with the discovery that the universe is expanding. For a long time quantum mechanics and cosmology developed  independently of one another. Yet the very discovery of the expansion would eventually draw the two subjects together because it implied the big bang where quantum mechanics was important for cosmology and for understanding and predicting our observations of the universe today. 
Textbook (Copenhagen)  formulations of quantum mechanics are inadequate for cosmology for at least four reasons: 1) They predict  the outcomes of measurements made by observers. But in the very early  universe no measurements were being made and no observers  were around to make them. 2) Observers were outside of the system being measured.  But we are interested in a theory of the whole universe where everything, including observers, are inside. 3) Copenhagen quantum mechanics could not retrodict the past. But retrodicting the past  to understand how the universe began is the main  task of cosmology. 4) Copenhagen quantum mechanics  required a  fixed classical spacetime geometry not least to give meaning to the  time in the Schr\"odinger equation. But in the very early universe spacetime is fluctuating quantum mechanically (quantum gravity) and without definite value. 
A formulation of quantum mechanics general enough for cosmology was started by Everett and developed by many. 
That effort has given us a more general  framework that is adequate for cosmology --- decoherent (or consistent) histories quantum theory in the context of semiclassical quantum gravity.  Copenhagen quantum theory is an approximation to this more general quantum framework that is appropriate for measurement situations.  We discuss whether further generalization may still be required.
\end{abstract}




\maketitle

\bibliographystyle{unsrt}

\tableofcontents


\section{Introduction}
\label{intro}
In the late '20s and early '30s of the last century quantum mechanics was revolutionizing our understanding of the physics of the very small. At roughly the same time another revolution in physics was beginning  that was to change our understanding of the physics of the very large.  This started with the discovery that the universe was expanding through the work of Lema\^itre, Hubble, and Slipher. For many years these two revolutions proceeded independently of one another. But in more recent times each has had  a significant impact on the other. 

Quantum mechanics is central to an understanding of important physical processes that take place in the very early universe. Well known examples are big bang nucleosynthesis, recombination, and, most importantly, the generation of quantum fluctuations in  matter density that grow under gravitational attraction to produce the large scale structure we see today in the distribution of the galaxies and in the temperature fluctuations of the cosmic background radiation (CMB). 

What is perhaps less widely appreciated that cosmology had a significant impact on our understanding of quantum mechanics. That is because the Copenhagen formulations that appear in many textbooks  are inadequate for cosmology for several reasons.  Quantum mechanics had to be generalized to apply to cosmology, and those generalizations led to a different understanding of the subject and what it is about.

This essay will describe very briefly the impact of cosmology on quantum mechanics and a little of the more general formulations of quantum theory that resulted. Throughout we attempt to explain the basic ideas with a minimum of technical development.\footnote{That is with a minimum of equations. For those who want the equations there are a large number of references, many by the author, that offer different levels of explanation.}
We trade lack of precision for hope of being more widely understandable. 

We should also stress that we will not address issues of the interpretation of quantum theory. Rather, we aim at describing how cosmology has affected what has to be interpreted. 

We begin in Section \ref{copenhagen} by describing  the status of  Copenhagen quantum mechanics (CQM) today. Section \ref{need} explains why this Copenhagen quantum theory is  inadequate for cosmology and why a generalization of it  is needed.  Section \ref{model} describes a model quantum universe in a very large box.  Section \ref{DH}  describes the decoherent (or consistent) histories generalization of Copenhagen quantum theory that is adequate for closed systems like the universe.
Section \ref{spacetime} describes further generalizations to deal with quantum spacetime geometry.
Section \ref{firstperson} describes how to get probabilities for the results of our observations from probabilities for histories of the universe. 
Section \ref{coprecovered} recovers  Copenhagen quantum mechanics  as approximation appropriate for measurement situations in more general formulations of quantum mechanics applicable to the universe. 
Section \ref{emergent}  describes how various formulations of quantum theory can be regarded as features of the universe that emerge along with other features that are a prerequisite to the formulations. 
Section 
Section \ref{conclusion} concludes by speculating on whether further generalizations will be required .


\section{The Copenhagen Quantum Mechanics  of  Laboratory Measurements (CQM)}
\label{copenhagen}
The Copenhagen quantum mechanics (CQM) found in standard textbooks\footnote{By `Copenhagen quantum mechanics', or `textbook quantum mechanics'  we mean the standard formulation that is found in many textbooks including the assumption of the quasiclassical realm of every day experience as described in \ref{qcrealm}.  We do not necessarily mean that it conforms exactly to what the founders of quantum mechanics meant by the `Copenhagen interpretation of quantum mechanics' \cite{Jam66}.} is arguably the most successful theoretical framework in the history of physics. It is central to our understanding of a vast range of physical phenomena, including atoms, molecules, chemistry, the solid state, how stars are formed, shine, evolve, and die, nuclear energy, thermonuclear explosions, how transistors work, and many, many, many, more phenomena. It has often been claimed that quantum mechanics is responsible for a significant fraction of the US GDP. 

\begin{figure}[t]
\includegraphics[width=6.5in]{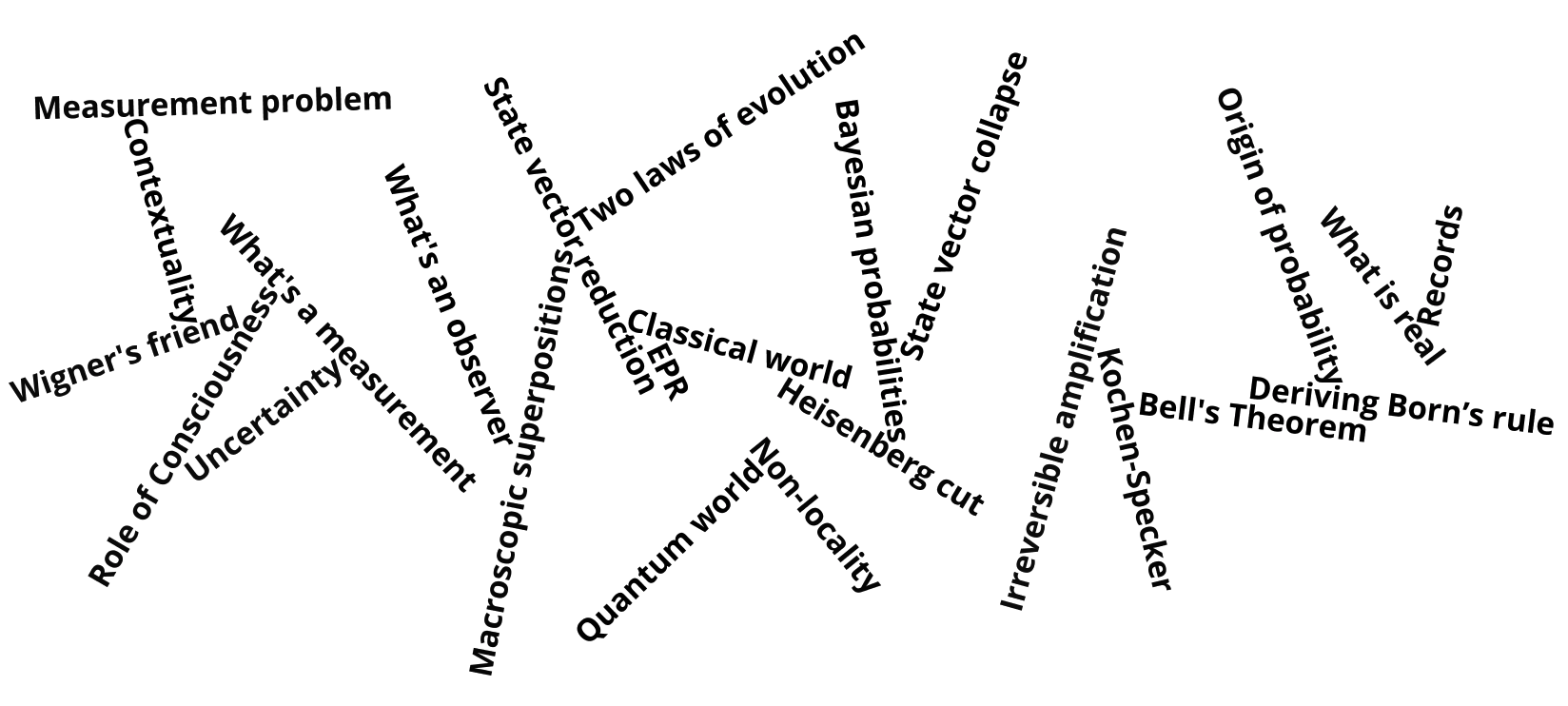}  
\caption{Foundational issues sometimes raised for Copenhagen quantum mechanics. None of these issues are discussed directly in this essay. Rather we aim to go beyond them with more general formulations of quantum theory. }
\label{cop-cloud}
\end{figure}

Work continues today to better understand Copenhagen quantum mechanics, to make its central notions of measurement and state vector reduction more precise, and to resolve `problems' that it is alleged to have like the `measurement problem' (Figure \ref{cop-cloud}). Even given these concerns, the author knows no mistake that was made because of them  in correctly applying Copenhagen quantum mechanics in the century since it was first formulated. 

 Despite its experimental success a  striking number of our most distinguished scientists are of the opinion that even the quantum mechanics of laboratory experiment will need to be revised (e.g \cite{qmmod}) for these concerns or perhaps others. 
 
None of these concerns will be  discussed here  directly. We do not aim to clarify or perfect Copenhagen quantum mechanics.  Rather we aim to move beyond it to  a quantum mechanics general enough for cosmology to  which Copenhagen quantum mechanics is  an approximation appropriate for laboratory situations.  The issues alluded to in Figure \ref{cop-cloud}  could have a very different form or be absent entirely in such a generalization (e.g. \cite{Sau10}). 

\section{Copenhagen Quantum Mechanics is Inadequate for Cosmology}
\label{need}
There are at least four reasons why textbook quantum mechanics is inadequate for cosmology. We describe them in this section.

\subsection{No  Quasiclassical Realm, No observers, No measurements, \\ in the Early Universe}
\label{qcrealm}
Copenhagen quantum mechanics assumes the quasiclassical realm of every day experience --- the wide range of time, place, and scale on which the deterministic laws of classical physics apply. Notable examples are the laws of gravity summarized by the Einstein equation and the laws of motion of continua summarized in the  Navier-Stokes equation. These and other classical laws are needed just to describe measurement apparatus and observers  and their operation in physical terms. Indeed, some formulations of Copenhagen quantum mechanics assumed the existence  entirely separate classical and quantum worlds with a kind of movable boundary between them (e.g. \cite{LL}). There is no evidence for such a division in cosmology. 

\subsection{Observers are Physical Systems within the Universe}
\label{obs-inside}
Copenhagen quantum mechanics predicts probabilities for the outcomes of measurements carried out on a subsystem of the universe  by another subsystem called the `observer' or `measuring apparatus' that is outside the subsystem being measured\footnote{ Individually and collectively observers are examples of the general concept of  IGUSes ---  information gathering and utilizing subsystems. It would be both more accurate and more accessible to present Copenhagen quantum mechanics in terms of IGUSes. But we aim at a discussion that is as broadly accessible as possible and to that end we will stick with the more broadly familiar if less nuanced `observer' in this paper.}.

But in the very early universe there were no measurements being made and no observers around to make them. Are we to believe that quantum mechanics does not apply to the early universe before the evolution of observers approximately 10 Gyr after the big bang?  Certainly not!  But then we need a generalization of Copenhagen quantum mechanics in which observers are physical systems {\it inside} the universe. We need a generalization of textbook quantum mechanics in  which observers and their measurements can be described in quantum mechanical terms  but play no special role in the theory's formulation and are not prerequisites for the theory's application\footnote{Sometimes one meets the suggestion that no such generalization is needed. The idea is that Copenhagen quantum mechanics  retrodicts the past when present  observers measure present  records of what went on in the past. But what exactly is a record? A record is a physical quantity today that is correlated with one or more past events with high probability. This probability is for a two time history of the events and the formation of a later record of them.  A generalization of Copenhagen quantum mechanics is needed to assign probabilities to such time histories and define what is meant by records.}.

\subsection{No Retrodiction}
\label{noretro}
In his now classic book {\it The Mathematical Foundations of Quantum Mechanics} \cite{vNeu32} John von Neumann formulated quantum evolution in terms of two laws. First, unitary evolution of the state vector described by the Schr\"odinger equation.  Second, the reduction of the state vector after an `ideal' measurement\footnote{vonNeumann had the opposite order of one and two but the one here is more used today} that disturbs the measured subsystem as little as possible. 

Starting from the state of an isolated subsystem at one time, and using these two laws, Copenhagen quantum theory can predict probabilities for the histories of later measurements an observer might decide to carry out on the same subsystem. This is the  sense in which Copenhagen quantum theory can predict the results of future measurements starting from the present state of the measured subsystem. 

Copenhagen quantum mechanics cannot retrodict the past  of a subsystem starting from its present quantum state. The Schr\"odinger differential equation can of course be evolved backward in time as well as forward.  But state vector reduction runs only one way --- forward in time, usually taken to be the direction in which entropy is increasing in the universe.  Thus retrodiction is impossible in Copenhagen quantum mechanics\footnote{The sole consideration of the past in Copenhagen quantum theory appears to be a paper by Einstein, Podolsky and Tolman \cite{EPT} which concluded `quantum mechanics must involve an uncertainty in the description of past events which is analogous to the uncertainty in the prediction of future events'. (Don Howard, private communication).}.

But retrodiction is central to cosmology. We use present data and and an initial quantum state to make a model of what went on in our past to simplify our predictions of the future \cite{Har98b}. We are interested in particular in a quantum framework that can can be used to construct a history of how the universe came to be the way it is today starting from its initial condition roughly 14Gyr ago using its quantum state.

\subsection{Fixed Classical Spacetime}
\label{fixed-spacetime}
Copenhagen quantum mechanics assumes a fixed spacetime geometry. This geometry defines the possible time directions for the `t' in the Schr\"odinger equation. States are defined on spacelike surfaces in this fixed spacetime and evolve  by the Schr\"odinger equation through a foliating family of such surfaces in this fixed spacetime. After an ideal measurement,  the state is `reduced' all across one of these spacelike surfaces that extend over the whole universe. Copenhagen quantum mechanics requires a fixed spacetime geometry just to make sense of these two modes of evolution.

But in the early universe near the big bang energy scales above the Planck scale $(\hbar c^5/G)^{1/2} \sim 10^{19} {\rm Gev}$ \ will be reached at which quantum gravity becomes important and spacetime geometry will fluctuate quantum mechanically and be without the definiteness required by Copenhagen quantum theory. 

Rather we expect Copenhagen quantum mechanics to emerge in the early universe {\it along with}  classical  spacetime geometry when the latter is sufficiently slowly varying  to provide the notions of time and spacelike surfaces that Copenhagen quantum mechanics assumes (e.g. \cite{har93c}).

The conclusion of the discussion in this section  is that CQM has to be generalized for quantum cosmology.  The rest of the paper discusses possible generalizations.


\section{A Model Quantum Universe in a Closed Box}
\label{model}

\begin{figure}[t]
\includegraphics[width=5in]{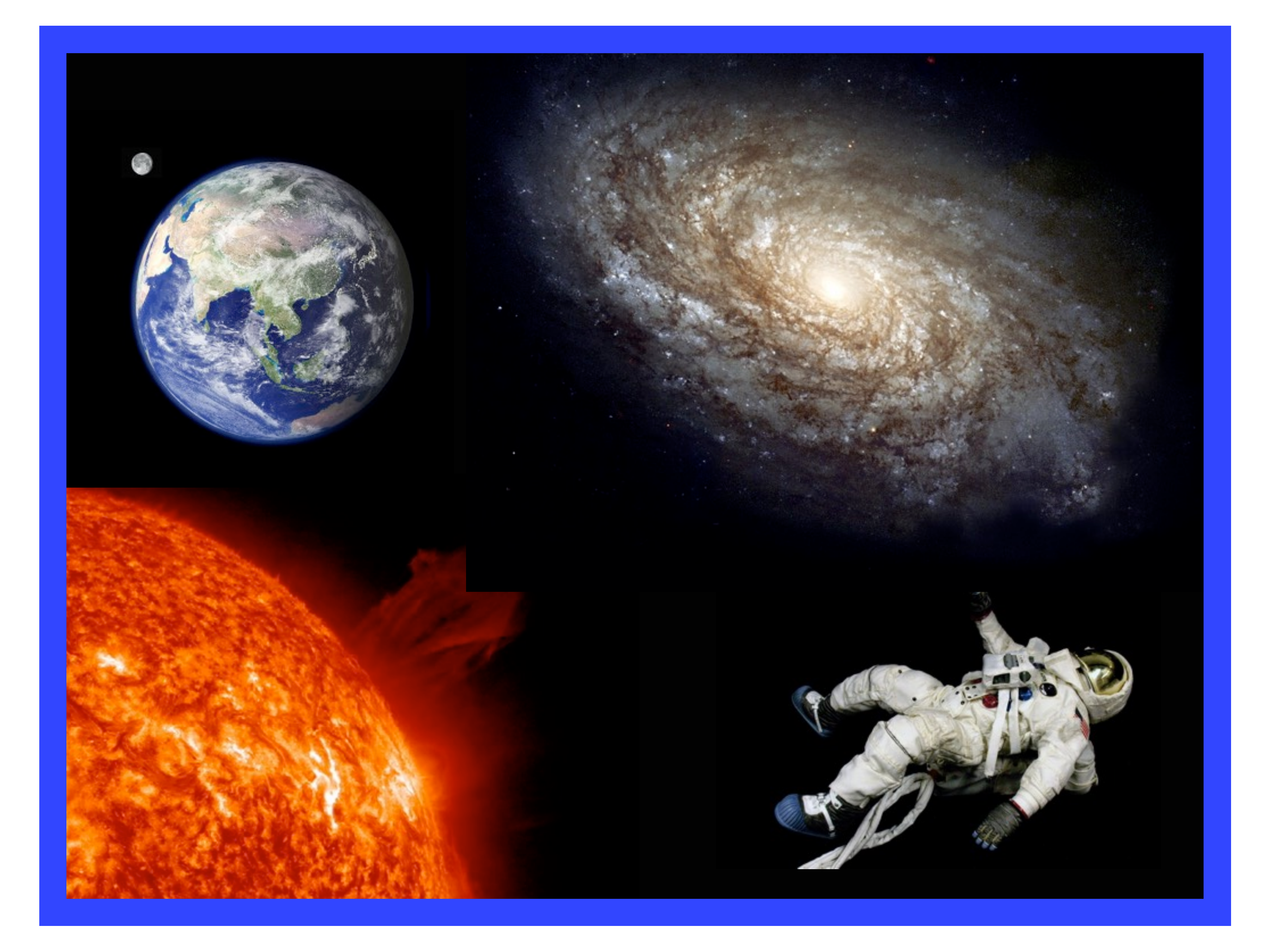}\hfill
\caption{A simple model of a closed quantum system is a universe of quantum matter fields inside a large closed box (say, 20,000 Mpc on a side) with fixed flat spacetime inside.  Everything is a physical system inside the box --- galaxies, stars, planets, human beings, observers and observed, subsystems that are  measured and subsystems that are measuring. The most general objectives for prediction are the probabilities of the individual members of decoherent sets of alternative coarse grained histories that describe what goes on in the box. That includes histories describing any measurements that take place there. There is no observation or other meddling with the inside from outside.  }
\label{box}
\end{figure}

To better understand how quantum mechanics works for the universe we first consider a model closed quantum system consisting of  a   large box say
 20,000 Mpc on a side, perhaps expanding, and containing particles and fields as suggested in  Figure \ref{box}. We  assume  a fixed, flat, background spacetime inside thus neglecting quantum gravity. This an excellent approximation in the realistic universe for times later than a very short interval  $\sim 10^{-43}sec.$ after the big bang. There is then a well defined notion of time in any particular Lorentz frame. The familiar apparatus of textbook quantum mechanics then applies ---  a Hilbert space, operators, states,  and their unitary evolution in time.  We assume a quantum field theory in the flat spacetime for dynamics. We will return briefly to what happens when quantum gravity is not neglected in Section \ref{spacetime}.

The important thing is that everything is contained within the box --- galaxies, planets, observers and
observed, measured subsystems, and any apparatus that measures them, you and me. There is nothing outside and no influence of the outside on the inside or the inside on the outside. 

The basic theoretical inputs for predicting what goes on in the box  are the  Hamiltonian $\H$ governing evolution and quantum state  $|\Psi\rangle$, written here in the Heisenberg picture for convenience, and assumed pure for simplicity. Input theory is  then \HP. The basic output are probabilities for what goes on in the box. More precisely the outputs of \HP\ are probabilities for the individual members of sets of coarse-grained alternative time histories of what goes on in the box. A few examples will make the important ideas introduced in this sentence  more concrete.

\subsection{Coarse-grained Histories in General}
\label{cghist} 

We might  be interested in a set of histories that describe the positions of the Moon in its motion around the Earth at a series of times $t_1,\cdots, t_n$.  We are then interested in the probabilities of the   alternative orbits that the  could Moon follow around the Earth.  Each orbit is an example of a {\it history} --- a sequence of events at a series of times. 
Relevant histories are {\it coarse-grained}  because we are only interested in positions defined to an accuracy consistent with our observations of position, and further  because positions are not specified at each and every time but only at a finite discrete sequence of times. Coarse grained histories can be said to {\it follow} certain variables and {\it ignore} others. In the present example the histories follow the center of mass of the moon and ignore variables that describe the interior of the Moon and Earth. In quantum mechanics there is no certainty that the coarse-grained history of the  Moon's center of mass  will follow a classical Keplerian orbit, but in the Moon's situation the probability predicted by \HP\ is vastly higher than for a non-classical orbit. 

Coarse graining is of the greatest importance in theoretical science, for example in connection with the meaning of regularity and randomness and of simplicity and complexity \cite{GH07}. Of course, it is also central to statistical mechanics, where a particular kind of coarse graining leads to the usual physico-chemical or thermodynamic entropy.  As we shall see, coarse graining  also plays a crucial role  in  the quantum mechanics of the universe, making possible the decoherence of alternative histories and enabling probabilities of those decoherent histories to be defined.

\subsection{Coarse Grained Histories of the Universe}
\label{cgcosmo}

In cosmology we are interested in the probabilities of sets of  certain coarse-grained alternative classical histories that describe the universe's expansion, the primordial nucleosynthesis of light elements, the formation and  evolution of the microwave background radiation, the formation  of the galaxies, stars, planets, the evolution of biota,  etc. Sets of alternative histories relevant for our observations in are highly coarse grained. They don't  describe everything that goes on in the universe --- every galaxy, star, planet, human history, etc., etc in all detail. Rather they follow  much coarser grained histories of the universe. And as we will see in quantum mechanics it is only histories that are sufficiently coarse grained for which the theory assigns probabilities at all!

In laboratory science we are interested in histories that describe the preparation, progress, and outcomes of a measurement situation.  In quantum cosmology there are no measurements of  the inside of the box by something outside it.  Laboratory measurements are described realistically,  as a correlation between one subsystem inside the box that includes the apparatus observers, etc and another subsystem inside that is thus measured. In this way measurements can be described in the quantum mechanics of the universe but play no
preferred role in the  formulation of quantum mechanics as they do in Copenhagen quantum theory.

The measuring apparatus and the orbit of the Moon are examples of systems that behave according to the laws of classical physics at a suitably coarse-grained level. Indeed most of our understanding of the universe on large scales is of its classical behavior. But classical behavior is not a given in a quantum universe. It is rather a matter of quantum probabilities.  A quantum system behaves classically when, in a suitably coarse-grained set of alternative histories, the probabilities are high for  for histories exhibiting correlations in time governed by deterministic classical laws (e.g. \cite{GH07,GH10,HHH08}).

\section{Decoherent Histories Quantum Theory (DH)}
\label{DH}

Decoherent histories  quantum mechanics (DH)  is a formulation of quantum theory that is general enough for cosmology. It has all the needed  generalizations of Copenhagen quantum mechanics  discussed in Section \ref{need}. 
It permits both prediction and retrodiction. Observers are physical systems within the closed universe. Measurements are physical processes within the universe. But neither observers nor their measurements play any preferred role in the formulation of DH.  As we will see, Copenhagen quantum mechanics  an approximation to this more general framework that is appropriate for measurement situations. 

Decoherent histories quantum theory (DH) is the work of many \cite{classicDH}, The formulation we present here is the work  Murray Gell-Mann and the author \cite{GH90}. On many essential points it coincides with the earlier independent consistent histories (CH) formulation of quantum theory of Giffiths and Omn\`es \cite{classicDH}.  DH  can be seen as a generalization, clarification, and, to some extent, a completion of the of the program started by Everett \cite{Eve57}. .

\begin{figure}[t]
\includegraphics[width=5in]{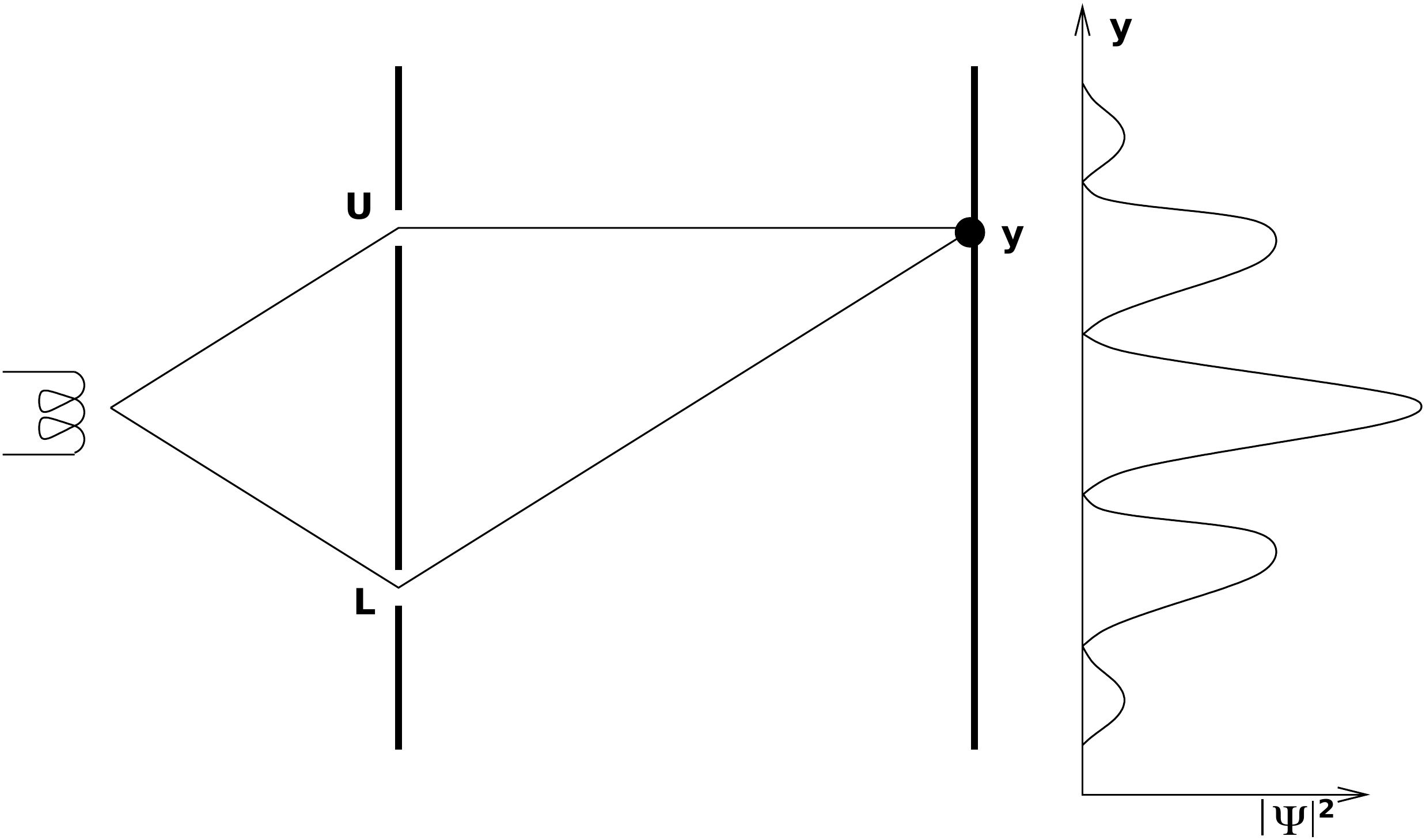}\hfill
\caption{The two-slit experiment.  An
electron gun at left emits an electron traveling towards a screen with
two slits, its progress in space recapitulating its evolution in time.  When
precise detections are made of an ensemble of such electrons at the
screen the classic interference pattern emerges --- evidence that the two histories interfere. 
is not possible, because of this quantum  interference, to assign a probability to the
alternatives of whether an individual electron went through the upper
slit or
the lower slit.  However, if the electron interacts with apparatus that
measures which slit it passed through, then these alternatives decohere
and
probabilities can be assigned.}
\label{twoslit}
\end{figure}

We will  not develop  the machinery necessary for an application of DH to cosmology in any detail here because we aim only  at an exposition of the principles behind the theory with a minimum of technical complication\footnote{ For a pedagogical introduction with more equations than here see, e.g. \cite{Har93a}.}. For definiteness we  first focus this discussion on the universe in a box of the previous section with a fixed, flat, spacetime geometry.  As described there,  we aim at predicting probabilities for the individual members of sets of coarse grained alternative histories of what goes on in the box.  But quantum interference is an  obstacle to assigning probabilities sets of alternative histories for any closed system.  Nowhere is this more simply illustrated than in the classic two-slit experiment shown in Fig. \ref{twoslit}.
\subsection{Two-Slit Experiment}
\label{twoslitamps}

In the two-slit setup shown in Figure \ref{twoslit} suppose an electron arrives at a position $Y$ on the detecting screen after passing through a screen with two slits $U$ and $L$. There is an obvious set of two alternative  histories of how it might have got to the detecting screen --- one in which it went through the upper slit ($U$) to arrive at the position interval $Y$ on the screen, and the other where it went through the lower slit ($L$) to arrive at $Y$. 
Quantum theory assigns amplitudes to the two individual histories set for example as in Feynman's sum-over-histories formulation of quantum theory \cite{FH64}. Denote   the quantum amplitudes  of the two histories  by $\Psi_U(Y)$ and $\Psi_U(Y)$.  The amplitude $\Psi(Y)$  to arrive at $Y$ is then the sum 
\be
\label{ampY}
\Psi(Y) =\Psi_U(Y)+\Psi_L(Y).
\ee

It would be inconsistent to assign probabilities to these histories that were squares of these amplitudes. The probability to arrive at $Y$ should be the sum of  the probabilities to arrive there by going through the upper or lower slit. But In quantum mechanics probabilities are squares of amplitudes and the sum of squares is not the square of the sum.
It would {\it inconsistent} with the standard rules of probability for a theory to predict probabilities for the two histories that were squares of amplitudes in this set. Specifically, 
 \begin{subequations}
\be
p(Y) \ne p_U(Y) +p_L(Y)
\ee
because
\be
\label{sumrule}
|\Psi(Y)|^2 = |\Psi_U(Y)+\psi_L(Y)|^2 \ne |\Psi_U(Y)|^2 +|\Psi_L(Y)|^2 .
\ee
\end{subequations}

Any formulation of quantum mechanics must therefore include a rule to specify which sets of alternative histories can be consistently assigned probabilities and which cannot.  In Copenhagen quantum mechanics the rule was simple:  Probabilities are assigned  only to sets of alternative histories that have been {\it measured} and not otherwise. If we measured which slit the electron when through then the interference pattern would be destroyed, the sum rule obeyed, and the probabilities consistent with the usual rules.

However, as discussed in Section \ref{need} we can't have such a rule in a quantum theory of cosmology which seeks to describe early  universe cosmology when no measurements were being made and there were no observers around to carry them out.   The more general rule that defines DH  is just this:
{\it  Probabilities can be assigned to just those sets of alternative histories of a closed system for which there is negligible interference between the individual histories in the set as a consequence of the Hamiltonian $(\H) $and the  state $(\Psi)$  that the closed system has.}  These probabilities are consistent with the rules of probability theory as a result of the absence of quantum  interference. Such a set of histories for which the mutual interference between any pair of histories is negligible  is said to {\it decohere.}

\subsection{Calculating Amplitudes for Histories in the Two-Slit Experiment} 
\label{amplitudes}
We can get an idea of how to calculate the amplitudes for histories  in the two-slit experiment just from the Schr\"odinger equation defining the quantum evolution of the electron's wave function. To keep the discussion simple we ignore the spin of the electron and consider only its position.   The state of the electron in the two-slit experiment  can then be described by a wave function of the form 
\be
\Psi=\Psi(x,y,t)
\label{wvfn}
\ee
using a coordinate $x$ for the horizontal direction and $y$ for the vertical direction in Figure \ref{twoslit} and assuming symmetry in the perpendicular direction.
   
Denote the initial state at time $t_0$  by $\Psi(x,y,t_0)$. This is a product of wave packet in the $x$-direction $\phi(x,t_0)$  and a wave function $\psi_0(y,t_0)$ localized at the gun, viz.
\be
\label{fullstate}
\Psi(x,y,t_0) =\psi(y,t_0)\phi(x,t_0) .
\ee
This wave function evolves in time by the Schr\"odinger equation
\be
\label{schrod}
i \hbar \frac{\partial \Psi}{\partial t} = H \Psi.
\ee
where $H$ is the Hamiltonian of a free particle interacting with the screens. 
We assume that   $\phi(x,t)$ is a narrow wave packet peaked to the left of the slits but moving to the right so as to reach the slits at time $t_s$ and the detecting screen at $t_d$. Thus, its progress in $x$ recapitulates evolution in time. 
After passing through the slits the wave function has the approximate form 
\begin{subequations}
\label{psiU}
\begin{align}
\Psi(x,y,t)&=\psi_U(y,t)\phi(x,t)  + \Psi_L(y,t)\phi(x,t),  \quad t_s<t<t_d . \label{PSUa} \\
                 &\equiv  \label{PSUc}  \Psi_U(x,y,t) + \Psi_L(x,y,t).
\end{align}
\end{subequations}
Here, in the first term, $\psi_U(y,t)$ is localized near the upper slit at time $t_s$ and spreads over a larger region of $y$ by the time $t_d$ that the wave packet hits the detecting screen. Similarly for the second term. When evaluated at the arrival position $x_d$ and time $t_d$ ,and projected on the interval of arrival $Y$, these are the amplitudes $\Psi_U(Y)$ and $\Psi_L(Y)$ in \eqref{ampY}.  They are wave functions of two branch state vectors $|\Psi_U(Y)\rangle$  and $|\Psi_L(Y)\rangle$ for the two histories in the set. The set does not decohere because 
\be
\label{decoh}
\langle\Psi_U(Y) | \Psi_L(Y)\rangle  \not\approx  0 
\ee  
so that the decoherence condition is not satisfied. The two histories interfere. 
 There will  be no consistent set of probabilities for this set of histories .

\subsection{A Simple Model of Decoherence}
\label{model-decoh}
To see how decoherence { can} occur suppose that near the slits there is a gas of particles (e.g. photons) that scatter off the electron but  weakly enough not to disturb its trajectory.  Even if the collisions do not affect the trajectories of the
electrons very much they can still carry away the phase correlations
between
the two histories.  A coarse graining that described
only
of these two alternative histories of the electron would then approximately
decohere as a consequence of the interactions with the gas given
adequate
density, cross-section, etc.  Interference is destroyed and consistent 
probabilities can be
assigned to these alternative histories of the electron in a way that
they
could not be if the gas were absent  ({\it cf.} Fig. 1).  The lost phase
information is still available in correlations between states of the gas
and states of the electron.  The alternative histories of the electron would
not decohere in a coarse graining that included both the histories of the
electron
{\it and} operators that were sensitive to the correlations between the
electrons and the gas.

Functioning in this way to dissipate relative phases between different histories the gas is an example of an {\it environment} (e.g. \cite{JZ85,Zur03,GH10,GH13}) and the example illustrates the widely occurring mechanism of {\it environmental decoherence}.

\begin{figure}[t]
\includegraphics[width=5in]{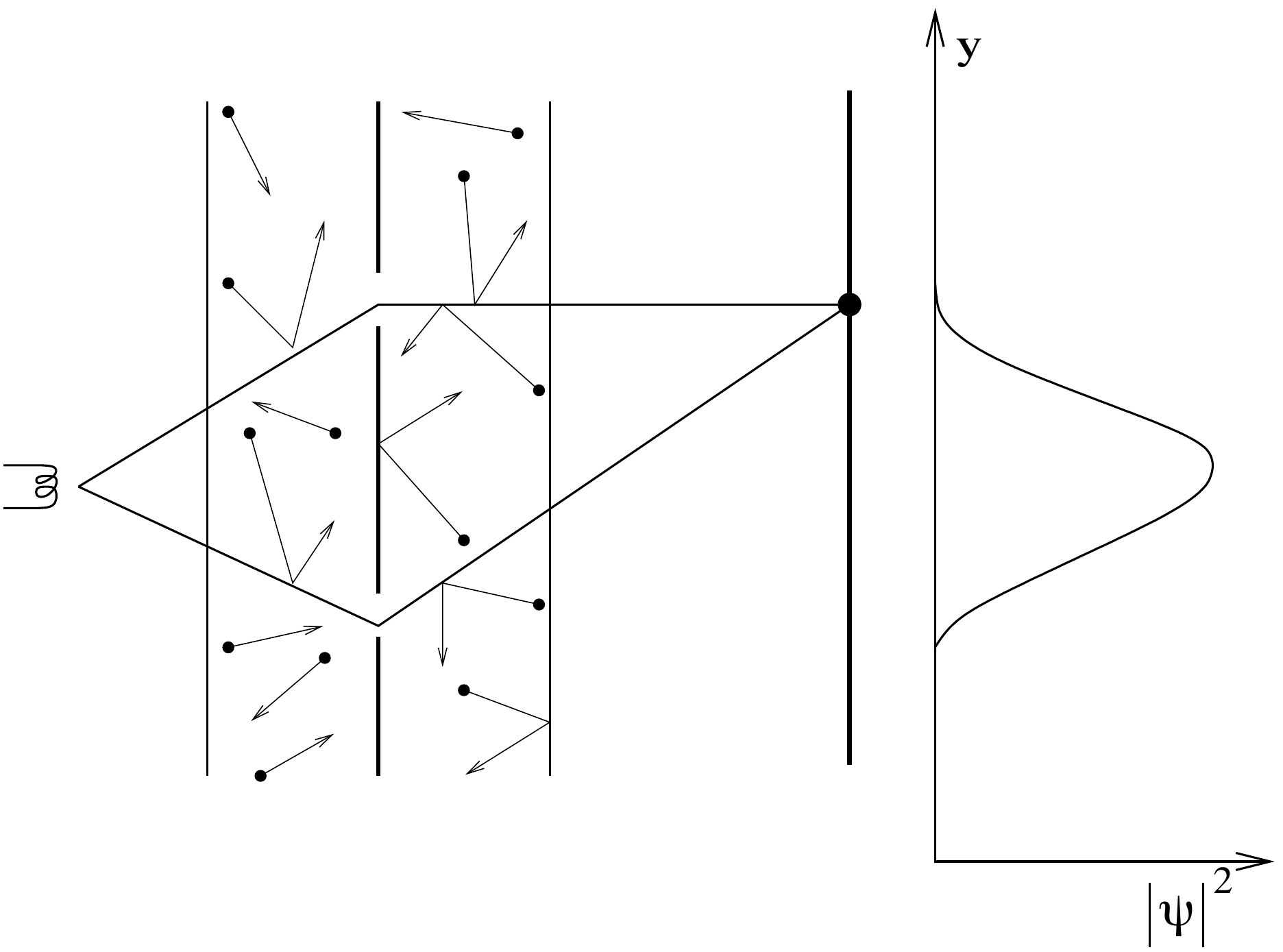}\hfill
\caption{The two-slit experiment
with an interacting gas. Near the slits light particles of a gas collide with
the electrons.  Even if the collisions do not affect the trajectories of the
electrons very much they can still carry away the phase correlations
between
the histories in which the electron arrived at point $y$ on the screen
by passing through the upper slit and that in which it arrived at the same
point
by passing through the lower slit.  A coarse graining that described
only
of these two alternative histories of the electron would approximately
decohere as a consequence of the interactions with the gas given
adequate
density, cross-section, etc.  Interference is destroyed and
probabilities can be
assigned to these alternative histories of the electron in a way that
they
could not be if the gas were not present ({\it cf.} Fig. 1).  .}
\label{twoslitgas}
\end{figure}

\begin{figure}[t]
\includegraphics[width=5in]{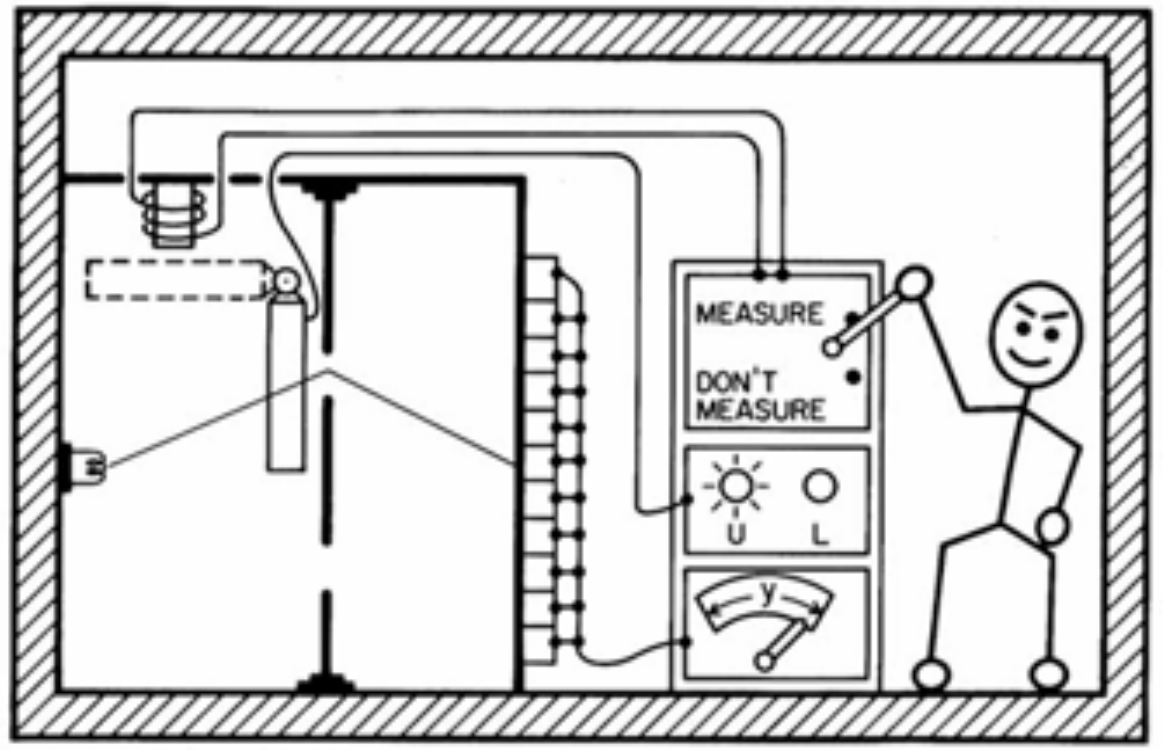}\hfill
\caption{A model closed quantum system containing an observer together
with the necessary apparatus for carrying out a two-slit experiment.
Alternatives for the system include 
whether the observer measured which slit the electron passed through or
did not, whether the electron passed through the upper or lower slit,
the alternative positions of arrival of the electron at the screen, the
alternative arrival positions registered by the apparatus, the
registration of these in the brain of the observer, etc., etc., etc. 
DH  assigns probabilities to the individual alternative
histories in such a set when there is negligible quantum mechanical
interference between them, that is, when the set of histories
decoheres.
}
\label{twoslitobs}
\end{figure}
 
The initial wave function  including the state of the gas is
\be
\label{fullstatewg}
\Psi(x,y,t_0) =\psi(y,t_0)\phi(x,t_0)\chi(t_0) , \quad t_0<t<t_s.
\ee
where $\chi(t_0)$ is the initial state of all the particles in the gas. After passing through the slits this becomes
\begin{subequations}
\label{psiUwgas}
\begin{align}
\Psi(x,y,t)&=\psi_U(y,t)\phi(x,t)\chi_U(t)  + \psi_L(y,t)\phi(x,t)\chi_L(t),  \quad t_s<t<t_d . \label{PSUga} \\
                 &\equiv  \label{PSUgc}  \Psi_U(x,y,t) + \Psi_L(x,y,t). 
\end{align}
\end{subequations}
where $\chi_U(t)$ and $\chi_L(t)$ denote the state of the gas particles that have scattered from the region of the upper and lower slit respectively. Then  $\Psi_U(x,y,t_d)$ and $\Psi_L(x, y,t_d)$ are the branch wave functions for the two histories that the electron arrived in the interval  $Y$ at time  $t_d$ after passing through either the upper or lower slit. 
The condition that this set of two histories decoheres is the absence of interference between these two branch wave functions, specifically
\be
\label{decohcond}
\langle \Psi_U(t_d)|\Psi_L (t_d) \rangle \approx 0 . 
\ee

In the simple two-slit model the set of alternative histories describing which slit ($U$ or $L$) the electron went through on its way to arrive at $Y$ on the further screen did not decohere as required by DH for predictions of probabilities cf. \eqref{decoh}. That was because the branch states representing the two histories were not orthogonal.  However,   the set of  two histories {\it does} decohere in this example because interactions with the gas dissipates relative phases. The decoherence condition \eqref{decohcond}  is satisfied because the state of the gas scattered from the upper slit $|\chi_U\rangle$ is orthogonal to the state of the gas scattered from the lower slit $|\chi_L\rangle$ if enough particles scatter. 

To see that gas ensures this orthogonality consider a single  gas particle $a$ whose initial state is $|\chi_a\rangle$ that scatters near the upper slit will be in a final state $S_U |\chi_a\rangle$ where $S_U$ is the S-matrix for scattering near the  upper slit. Similarly for the lower slit with $S_L|\chi_a\rangle$  If $N$ gas particles scatter the overlap between $|\chi_U\rangle $ and $|\chi_L\rangle$ will be proportional to  $N$  inner products of the form
\be
\label{overlap} 
\prod_{a=1}^N |\langle\chi_a| S_U^\dagger S_L|\chi_a\rangle|. 
\ee
Since the two final state vectors $S_U |\chi_a\rangle$ and $S_L|\chi_a\rangle$  are different each individual product has a magnitude less than $1$. The product of a very large number of such products will be near zero implying decoherence as in \eqref{decohcond}. 

\subsection{General Features of Quantum Mechanics Illustrated by the \\ Two-Slit Experiment}
\label{gen-two-slit}
This two-slit example illustrates two general features of the quantum mechanics of histories that are central to its formulation.  

First, it illustrates  that  sets of alternative histories must decohere in order to have probabilities consistent with the rules of probability theory. 

Second, the two-slit experiment illustrates the crucial role that coarse graining plays by making decoherence possible. Sets of fine-grained alternative histories like the set of all possible Feynman paths of a particle generally do not decohere except in trivial cases. Some degrees of freedom must be ignored to carry away the phase information between individual coarse grained histories that can be followed.  It is remarkable fact that in quantum mechanics some information must be lost to have any information at all.

Third, the two-slit experiment leads to understanding the central role that cosmology plays in implementing environmental decoherence. 
Mechanisms of decoherence are wide spread in our epoch of the  universe.  We see that cosmology is not just important for formulating quantum but also for implementing the decoherence that is necessary for its predictions.

\section{Generalized Decoherent Histories Quantum Mechanics of Quantum Spacetime (GDH)}
\label{spacetime}
The exposition of decoherent histories quantum mechanics of closed systems (DH)  in Section \ref{DH} assumed one fixed classical background spacetime. Histories describe the motion of particles and the evolution of fields {\it in} this background spacetime. But in a quantum theory of spacetime geometry we do not expect just one classical   spacetime geometry. Rather we expect an ensemble of possible classical spacetimes for which quantum theory  predicts probabilities for which we observe from the theories of dynamics and the quantum state\footnote{Theories of spacetime like general relativity are more straightforwardly summarized by an action than a Hamiltonisn but we will  denote the dynamical theory by $H$ however it is expressed.} $(\H,\Psi)$. 

It is conceptually simple but technically complicated to generalize the DH in Section \ref{DH} to include dynamical spacetime. Here we describe only the three simple conceptual changes from DH . We call the resulting generalization ``Generalized Consistent Histories Quantum Mechanics'' (GDH). The three important ingredients in GDH are:
\begin{itemize}
\item{\it Fine-grained histories:}. The set of possible alternative histories of spacetime geometry and matter field configurations  that our universe might have each specified by one spacetime geometry and one field history.
\item {\it Coarse grained histories:}   These are defined by a partition of the set of fine grained histories into exclusive bundles of fine-grained histories. The bundles are the coarse-grained histories.
\item {\it A measure of quantum interference} between different coarse-grained histories constructed by Feynman path integrals over $\int\exp\{-iS/\hbar\}$ over the fine-grained histories in the coarse grained sets. The  functional  $S$ is the action of the dynamical theory $H$ and the integral is weighted by a wave function of the universe representing $\Psi$  like the no-boundary quantum state \cite{HH83,HHH18}.
\end{itemize} 

\noindent Details can be found in   \cite{Har95c, HHH18}. So formulated GDH would predict probabilities for the individual members of decoherent   sets of alternative coarse-grained cosmological classical histories of geometries and fields \footnote{At the time of writing,  ways  of implementing coarse graining and decoherence for histories of spacetime geometry are at an early stage of development.}.

\section{Back to Observations, Measurements, Observers and Copenhagen Quantum Mechanics as an Approximation.}
\label{meas-obs}

\subsection{Predicting Probabilities for the Results of Observations}
\label{firstperson}
When implemented in a quantum framework $Q$ like DH or GDH a theory \HP\ predicts probabilities for which of a set of alternative coarse-grained histories of the universe occurs.   A triple $(H,\Psi,Q)$ is tested by its success in predicting the results of our observations of the universe. Predictions for our observations are through  probabilities for histories supplied by \HP\ conditioned on data $\Dobs$ describing our observational situation, including a description of any observers making the observation,  and assuming  here for simplicity that there is a unique instance of $\Dobs$ in the universe\footnote{In the vast universes contemplated by current theories of inflation $\Dobs$ is not likely to be  unique.  Then a further assumption is needed about which of the instances of $\Dobs$ is us (e.g. \cite{HH15b,HH06,HS07,SH10}}.

By way of example,  consider a prediction for the  temperature of  the CMB that will  be observed in a particular experiment. Locally the experiment would be described by data $\Dobs$ that would include of a description of the detector, where it is pointing, whether its shielded from radiation from the ground,  a description of the observers running the experiment, etc. But the theory \HP\  by itself does not supply a probability for the one temperature  observed. The temperature varies in time getting lower as the universe expands. Rather, \HP\ supplies probabilities for histories of how  the temperature varies over time.  But we don't observe entire four-dimensional histories of the CMB. We observe the CMB at a narrow range of times compared about 14Gyr after the big bang. Predictions for our observations of the CMB  thus depend on a  $\Dobs$ that  include the astronomical observations  that determine the time from the big bang that when  observation is being made --- the observed Hubble constant, the cosmological constant, and the mean mass density for instance.

\subsection{Copenhagen Quantum Mechanics Recovered as an Approximation}
\label{coprecovered}

 The construction, operation, and outcomes of  realistic  measurement situations can be described by an appropriate set of four-dimensional histories.
In a measurement situation a subsystem --- the apparatus  --- interacts over a short time with another subsystem, --- the measured subsystem ---  to produce records of the values of the measured quantities. DH supplies probabilities for the values of these records as the probabilities of appropriate histories.

Figure \ref{twoslitobs} shows a simple model of a measurement situation in which an observer is carrying out a two-slit experiment. The observer's apparatus measures which slit the electron went through, $U$ or $L$, and the position interval of arrival at the screen $Y$. The results $U$ or $L$ and $Y$ are recorded by the apparatus. All the ingredients for the application of Copenhagen quantum theory are there. There is a measured subsystem --- the electron. There is a measurement apparatus which records the outcomes, and there is an observer. Given the initial state of the electron at the gun the probabilities for these results can be calculated with Copenhagen quantum theory. 

A measurement situation like this might have been constructed somewhere in the history of the universe.  As such, the probabilities for the same recorded outcomes could be calculated with DH.  
When the CQM probabilities are close to those of DH  for all the records we have recovered CQM as an approximation to DH in suitable measurement situations. A fairly detailed discussion of how this works was given in \cite{Har91a}, Section II.10. Other measurement models that could be used to analyze this connection can be found in \cite{vNeu32,LB39,Wig63}.  Of course we expect that measurement situations will occur only at certain epochs of the universe ---  not in the early universe, not where there were no IGUSes to construct them, not  without classical spacetime, etc. These limitations are  reasons why we need a generalization of CGM for cosmology. 

\section{Emergent Formulations of Quantum Mechanics for Cosmology}
\label{emergent}
Sections \ref{copenhagen} to \ref{spacetime} described how the  Copenhagen quantum mechanics of measurement situations (CQM) can be generalized to provide a quantum framework for prediction in quantum cosmology in two steps. The first was the decoherent histories quantum mechanics (DH) of a closed system like the universe with a fixed background spacetime geometry. The second step is  a generalization of DH to a decoherent histories quantum mechanics {\it of} the spacetime geometries and quantum fields namely GDH. 

Each of these formulations assumes the existence particular features of our universe.  
Copenhagen quantum mechanics assumes  both a classical world containing laboratories with measuring apparatus and observers that obey the laws of classical physics and a quantum world of alternatives that are measured. Decoherent histories quantum mechanics assumed a fixed background spacetime geometry. These features are examples of excess baggage\footnote{In many advances in physics some previously accepted general idea was found to be unnecessary and dispensable. The idea was not truly a general feature of the world, but only perceived to be general because of our special place in the universe and the limited range of our experience. It was excess baggage which had to be jettisoned to reach a more a more general perspective \cite{Har90b}}. They are true physical facts but  describe special  situations in the the universe that occur only at particular  times and circumstances. We can say that they {\it emerged} at those particular times and places in the evolution of our universe.

For example,  in the early stages of the universe,  spacetime geometry was fluctuating quantum mechanically with no fixed value. As a consequence there was no fixed notion of time and  no fixed notion of spacelike or timelike. There then could  be no notion of quantum states  defined on a foliating family of spacelike surfaces and, perforce,  no notion of the unitary evolution of states through such families of surfaces. There were no isolated subsystems, and no notion of measurement of one by another. There were no observers to exploit the regularities of the universe.   The features of the universe assumed in the formulation of the Copenhagen quantum mechanics and DH were not there in the early universe. 
 
 Not much further on in the evolution of the universe classical spacetime emerged.  With it emerged the quasiclassical realm  -----  the wide range of time place and scale also on which the deterministic laws of classical physics hold \cite{Har11} .  With it also emerged the notions of time, states on spacelike surfaces and their unitary evolution by the Schr\"odinger equation \cite{har93c}.   
 Eventually, galaxies, stars, planets, and biota emerged $\sim 10^{20}$ years later and with them emerged  the quasiclassical realm, and  the observers, and measurements that are the subjects of Copenhagen quantum theory. It is then, and only then, that Copenhagen quantum mechanics is an appropriate approximation to the more general formulations\footnote{Thus only sufficiently late epochs of the universe will contain the observers necessary for Copenhagen quantum mechanics. But also we cannot expect observers (including ourselves) very late in the universe's history. The cosmological constant causes the universe to expand and cool exponentially quickly. Stars exhaust their thermonuclear fuel and die out. Black holes evaporate. The density of matter and the temperature approach zero. Mechanisms of environmental decoherence fail \cite{HHff}. The universe becomes cold, dark and inhospitable to the formation and functioning of observers \cite{har16a}.}.

\section{Conclusion --- Are We Finished Generalizing Quantum Mechanics?}
\label{conclusion}

The impacts of cosmology on quantum mechanics are at least these:  

$\bullet$\ Cosmology requires a quantum mechanics of a closed system in which measurements and observers can be described but are not central the formulation of the theory.

$\bullet$\ Cosmology requires a formulation of quantum mechanics general enough to deal with quantum spacetime geometry, the big bang, and what emerged from it, 

$\bullet$\  Cosmology requires a formulation of quantum mechanics that can describe the emergence of classical spacetime in the early early universe and along with it the quasiclassical realm --- the wide range of time, scale, and place that the classical laws of physics apply.

$\bullet$\ Cosmology requires a formulation of quantum mechanics to which the Copenhagen quantum mechanics is an approximation appropriate for measurement situations.

GDH potentially provides a generalization of  CQM that  can  be applied to the prediction of the outcomes of our cosmological observations from theories of the universe's quantum state $\Psi$ and dynamics $\H$. GDH  is therefore a formulation that is adequate for quantum cosmology. But it seems unlikely that it is the only possible adequate formulation.  It is a large and daring extrapolation to go from a quantum mechanics of laboratory situations to a quantum mechanics of the whole universe.  This is especially the case because the data testing the extrapolation is very limited.  Generalizations different from GDH are therefore likely to be possible.  Indeed, different routes to  generalizations have already been suggested \cite{Har04,Har07,Har08,GH11,Har16,SG04}. 
A generalization of CQM does not just consist of a theory of dynamics $(H)$ plus a theory of the quantum state $(\Psi)$.  It also must specify the quantum framework for prediction $(Q)$. For the input theory in quantum cosmology we should write $(H,\Psi,Q)$. 

We hope to find  generalizations $Q$ that have less excess baggage than GDH such as a principle of superposition for quantum states of the whole universe that can never be realized  experimentally.  Perhaps even the notion of spacetime geometry is excess baggage.  We hope to find generalizations that are more deeply connected to ideas in contemporary fundamental physics such as duality (e.g \cite{HH12}).  We hope to find generalizations that more deeply unify the the theory  of the quantum state $\Psi$ and the theory of dynamics $\H$.  But above all we hope to find generalizations that lead to definitive tests either in experiments in the laboratory or by large scale observations.  {\it We are not yet finished understanding the impact of cosmology on quantum mechanics. }

Cosmology, quantum mechanics, and fundamental physics are inextricably but fruitfully liked. The big bang is the place in the universe where the near Planck energy scales that characterize much of today's fundamental theory are reached. The observable consequences of the big bang by which these theories might be tested are scattered over the large scales of space and time observable today.  We should not imagine that predicting these consequences from $(\H,\Psi)$ is simply an a matter of calculating in a fixed quantum framework established by low energy experiment. 
Rather, as we have seen in this paper, our understanding of cosmology impacts our understanding of what is the quantum mechanics that applies to the universe as a whole. 

We end with a quotation that is  the final paragraph of  the first paper on decoherent histories quantum mechanics \cite{GH90}:

``We conclude that resolution of the problems of interpretation presented
by quantum mechanics is not to be accomplished by further intense scrutiny of
the subject as it applies to
reproducible laboratory situations, but rather through an
examination of the origin of the universe and its subsequent history.
Quantum mechanics is best and most fundamentally understood in the context
of quantum cosmology.  The founders of quantum mechanics were right
in pointing out that something external to the framework of wave function
and Schr\"odinger equation {\it is} needed to interpret the theory.  But
it is not a postulated classical world to which quantum mechanics does not
apply.  Rather it is the initial condition of the universe that, together
with the action function of the elementary particles and the throws of quantum
dice since the beginning, explains the origin of quasiclassical domain(s)
within quantum theory itself''.

\acknowledgments  

The author thanks Murray Gell-Mann, Thomas Hertog, and Mark Srednicki  for discussions of the quantum mechanics of the universe over a long period of time. He thanks the Santa Fe Institute for supporting many productive visits there. He thanks the organizers of the conference 90 Years of Quantum Mechanics in Singapore which was the occasion of a talk on which this work is partially based. The this work was supported in part by the National Science Foundation under grant PHY15-04541 and PHY18-18018105 .



\end{document}